The Law of Conservation of Energy in Chemical Reactions

I. A. Stepanov

Latvian University, Rainis bulv. 19, Riga, LV-1586, Latvia

## Abstract

Earlier it has been supposed that the law of conservation of energy in chemical reactions has the following form:

$$\Delta U=\Delta Q-P\Delta V+\sum_i \mu_i\Delta N_i$$

In the present paper it has been easy proved by means of the theory of ordinary differential equations that in the biggest part of the chemical reactions it must have the following form:

$$\Delta U=\Delta Q+P\Delta V+\sum_i \mu_i\Delta N_i$$

The result obtained allows to explain a paradox in chemical thermodynamics: the heat of chemical processes measured by calorimetry and by the Van't-Hoff equation differs very much from each other. The result is confirmed by many experiments.

## I. Introduction

The law of conservation of energy for heat exchange is the following one:

$$\delta Q=\Delta U+\delta A \qquad (1)$$

Here $\delta Q$ is the heat introduced to the system, $\Delta U$ is the change in the internal energy and $\delta A$ is work done by the system. Further it will be assumed that $\delta A=P\Delta V$.

For chemical processes this law is written in the following form [1]:

$$\Delta U=\Delta Q-P\Delta V+\sum_i \mu_i \Delta N_i \qquad (2)$$

where $\mu_i$ are chemical potentials and $\Delta N_i$ are the changes in the number of moles. It is due to the fact that the motive force of heat exchange is the heat introduced to the system but a motive force of a chemical reaction is the change in the internal energy. In this paper it is assumed that $\Delta Q$ is negative for exothermic reactions.

Using the theory of differential equations of physical processes it is possible easy to prove that the energy balance in the form of (2) for the biggest part of the chemical reactions is not correct. In the biggest part of the chemical reactions the law of conservation of energy must have the following form:

$$\Delta U=\Delta Q+P\Delta V+\sum_i \mu_i \Delta N_i \qquad (3)$$

## II. Theory

Pay attention that if P=const and A is the work of expansion then dQ and dA=PdV are exact differentials. Let's consider heat exchange, one introduces the quantity of heat $\Delta Q$ in the system ($\Delta V=0$): $\Delta Q=\Delta U$. Now let's suppose $\Delta V>0$. Let's find out, is it necessary to write $\Delta Q=\Delta U-P\Delta V$ or $\Delta Q=\Delta U+P\Delta V$. It is necessary to add the term $P\Delta V$ to the right side of the equation. If $\Delta Q>0$ then $\Delta U>0$ and $P\Delta V>0$. Then, according to the theory of differential equations of physical processes one gets (1) with $\Delta A=P\Delta V$.

Here a traditional method from differential equations was used [2]. If there is an equality $df=\pm\alpha dx\pm\beta dy\pm\gamma dz$ ($\alpha$, $\beta$ and $\gamma>0$) and one does not know what signs must there be in front of $\alpha$, $\beta$ and $\gamma$ one proceeds as following. Let dy and dz=0. Let df<0, if dx<0 then $+\alpha$. The signs before $\beta$ and $\gamma$ are found analogically.

Let's consider an exothermic and an isothermic reaction. Let's suppose that $\Delta V=0$. In this occasion the 1st law of thermodynamics will be the following one:

$$\Delta U=\Delta Q+\sum_i \mu_i \Delta N_i \tag{4}$$

Now let's suppose that $\Delta V\neq 0$. For the biggest part of the exothermic reactions with $\Delta V\neq 0$ the difference in the volume $\Delta V$ is less than zero (in isothermal case), for example, the following reaction:

$$2H_2+O_2=2H_2O \tag{5}$$

The term $P\Delta V$ must be added to the right side of (4). Let's analyze what sign must it have. $\Delta U$ is less than 0, $P\Delta V$ is less than 0. Hence the 1st law of thermodynamics for a chemical reaction will have the following form:

$$\Delta U=\Delta Q+P\Delta V+\sum_i \mu_i \Delta N_i \tag{6}$$

Really, $dU=dU_1+dU_2+dU_3$ where $dU_1=dQ$, $dU_2=\pm PdV$ and $dU_3=\sum \mu_i dN_i$. If $dU>0$ then $dU_1$, $dU_2$ and $dU_3>0$ and $dV>0$ whence the sign before P is plus.

It is important to stress that for heat exchange $(\partial U/\partial V)_S<0$ but for an endothermic reaction $(\partial U/\partial V)_{S,N}>0$. In an endothermic reaction one heats the reactants, dU increases and dV increases. Equation (6) describes this process correct but according to (2) the volume must decrease. If an open system with no chemical reactions expands on its own then its internal energy is exausted and $(\partial U/\partial V)_{S,N}<0$.

## III. Discussion and Experimental Checking

The 1st law of thermodynamics in the form of (6) was obtained not strictly in [3]. In the present work a strict derivation is given.

Using the result obtained in the present paper it is possible to explain a paradox in chemical thermodynamics: the heat of chemical reactions, that of dilution of liquids and that of micelle formation measured by calorimetry and by the Van't-Hoff equation differ significantly [4-6]. The difference is far beyond the error limits. According to thermodynamics the Van't-Hoff equation must give the same results as calorimetry  because the Van't-Hoff equation is derived from the first and the second law of thermodynamics without simplifications. Nobody succeeded to explain this paradox. Most probably, the reason is that in the derivation of the Van't-Hoff equation it is necessary to take into account the law of conservation in the form of (6), not of (2).

If to derive the Van't-Hoff equation taking into account (6) the result will be the following one:

$$d/dT \ln K=\Delta H^{*0}/RT^2 \qquad (7)$$

where K is the reaction equilibrium constant and $\Delta H^{*0}=\Delta Q^0+P\Delta V^0$. Let's check experimentally the result obtained.

In [7] the dependence of $N_2O_4=2NO_2$ reaction equilibrium constant on the temperature was given which was found experimentally (Table1). From Table 1 one sees that the equilibrium constant at 293, 303 and 323 K obeys the equation

$$\ln K = 22{,}151-59677/RT \qquad (8)$$

with high accuracy. Values calculated by (8) differ from experimental ones at  0,1;  0,2 and 0,03%, respectively. The equilibrium constant at 273 K deflects from this relationship. Dependence of  K on temperature is a broken line. The points 293, 303 and 323 K are at one segment, 273 K is at the other one.  From Eqs. (7) and (8) it is possible to calculate $\Delta H^{*0}$ and $\Delta Q^0=\Delta H^{*0}- P\Delta V^0$. We suppose that gas is ideal. These values are given in Table 1. There also the values of the heat of reaction ΔQ are given calculated according to [8]. One sees a good agreement with the present theory. Pay attention that K in (8) is the true reaction equilibrium constant which depends on the activities.

In [9] the following reaction was considered:

$$2Zn(gas)+Se_2(gas)=2ZnSe(solid) \quad (9)$$

$$\Delta G_f^0=-723{,}614+382{,}97T \text{ kJ/mol, } 1260<T<1410 \text{ K} \quad (10)$$

As ln K=-$\Delta G^0$/RT, from (10) one may find $\Delta H^{*0}$ and $\Delta Q^0=\Delta H^{*0}-P\Delta V^0$. The heat of reaction ΔQ is taken from [10] for T=1300 (Table 2). For T=1400 the data in [10] are absent.

In [11] the following reaction was given:

$$2HgTe(solid)=2Hg(gas)+Te_2(gas) \quad (11)$$

$$\ln K=-348496/RT+13{,}77;\ 778<T<943 \text{ K} \quad (12)$$

The heat of this reaction is given in Table 2.

In [12] the following reaction was given:

$$2ZnS(solid)=2Zn(gas)+S_2(gas) \quad (13)$$

$$\ln K=-774142/RT+21{,}976;\ 1095<T<1435 \text{ K} \quad (14)$$

The heat of this reaction is given in Table 2.

The equilibrium constant K in Eqs. (12), (14) depends on pressure, not on fugacity. But at a temperature many hundreds grades Kelvin and at the atmospheric pressure one can neglect intermolecular interactions and pressure is very close to fugacity.

In [13] the following reaction was considered:

$$2Te(liquid)=Te_2(gas) \quad (15)$$

$$\ln P=-119{,}755/RT+11{,}672469, \quad 722{,}65<T\leq 800 \text{ K} \quad (16)$$

$$-116{,}706/RT+11{,}21345, \quad 800<T<921{,}6 \text{ K}$$

$$-114{,}045/RT+10{,}8661, \quad 921{,}6<T<1142 \text{ K.}$$

Assuming K=P one can build Table 3.

One sees that experiment is in a good agreement with the theory developed in the present paper.

Table 1

Dependence of the heat of $N_2O_4$=2$NO_2$ reaction on the temperature [7]

| **T,°K** | **K (atm) [7]** | **$\Delta H^{*0}$ (kJ/mol)** | **$\Delta H^{*0}$-$P\Delta V^0$ (kJ/mol)** | **$\Delta Q$ (kJ/mol) [8]** |
|---|---|---|---|---|
| **273,15** | 0,01436 | - | - | - |
| **293,15** | 0,09600 | 59,68 | 57,24 | 57,30 |
| **303,15** | 0,2140 | 59,68 | 57,16 | 57,25 |
| **323,15** | 0,9302 | 59,68 | 56,99 | 57,15 |

Table 2

The heat of some chemical reactions measured by the Van't-Hoff equation and by calorimetry

| **$2Zn(gas)+Se_2(gas)=2ZnSe(solid)$ [9]** | | | |
|---|---|---|---|
| **T,°K** | **$\Delta H^{*0}$ (kJ/mol) (10)** | **$\Delta H^{*0}$-$P\Delta V^0$ (kJ/mol)** | **$\Delta Q$ (kJ/mol) [10]** |
| 1300 | -723,614 | -691,20 | -690,80 |
| **$2HgTe(solid)=2Hg(gas)+Te_2(gas)$ [11]** | | | |
| **T,°K** | **$\Delta H^{*0}$ (kJ/mol) (12)** | **$\Delta H^{*0}$-$P\Delta V^0$ (kJ/mol)** | **$\Delta Q$ (kJ/mol) [10]** |
| 800 | 348,496 | 328,56 | 329,50 |
| 900 | 348,496 | 326,06 | 325,72 |
| **$2ZnS(solid)=2Zn(gas)+S_2(gas)$ [12]** | | | |
| **T,°K** | **$\Delta H^{*0}$ (kJ/mol) (14)** | **$\Delta H^{*0}$-$P\Delta V^0$ (kJ/mol)** | **$\Delta Q$ (kJ/mol) [10]** |
| 1100 | 774,142 | 746,72 | 753,00 |
| 1200 | 774,142 | 744,22 | 750,00 |
| 1300 | 774,142 | 741,74 | 746,92 |
| 1400 | 774,142 | 739,24 | 743,76 |

Table 3

The heat of 2Te(liquid)=$Te_2$(gas) reaction measured by the Van't-Hoff equation [13] and by calorimetry

| **T,°K** | **$\Delta H^{*0}$ (kJ/mol) (16)** | **$\Delta H^{*0}-P\Delta V^0$ (kJ/mol)** | **$\Delta Q$ (kJ/mol) [10]** |
|---|---|---|---|
| **800** | 119,76 | 113,11 | 112,92 |
| **900** | 116,71 | 109,27 | 109,41 |
| **1000** | 114,05 | 105,74 | 105,97 |
| **1100** | 114,05 | 104,91 | 102,59 |